\documentclass[a4paper,11pt]{article}
\usepackage{aaskaiid}
\usepackage{aas_macros}%
\usepackage{orcidlink}
\title{Solar Radio Burst Fine Structures}
\ShortTitle{Solar radio burst fine structures} 
\ShortName{Kontar et al.} 

\author[1]{Eduard P. Kontar \orcidlink{0000-0002-8078-0902}}
\author[1]{Daniel Clarkson \orcidlink{0000-0003-1967-5078}}
\author[2]{Hamish Reid \orcidlink{0000-0002-6287-3494}}
\author[1]{Yingjie Luo \orcidlink{0000-0002-5431-545X}}
\author[3,1]{Nicolina Chrysaphi \orcidlink{0000-0002-4389-5540}}
\author[4]{Alexey Kuznetsov \orcidlink{0000-0001-8644-8372}}
\author[5,6,1]{Galina Motorina \orcidlink{0000-0001-7856-084X}}
\author[7]{Carine Briand \orcidlink{0000-0003-0271-9839}}

\affiliation[1]{School of Physics \& Astronomy, University of Glasgow, G12 8QQ, UK}
\emailAdd{Daniel.Clarkson@glasgow.ac.uk}

\affiliation[2]{Mullard Space Science Laboratory, UCL, London, UK}
\emailAdd{hamish.reid@ucl.ac.uk}

\affiliation[3]{INAF-IAPS, Via del Fosso del Cavaliere, 100, 00133 Rome, Italy}
\emailAdd{nicolina.chrysaphi@inaf.it}

\affiliation[4]{Institute of Solar-Terrestrial Physics, Irkutsk 664033, Russia}
\affiliation[5]{Central Astronomical Observatory at Pulkovo of Russian Academy of Sciences, St. Petersburg, 196140, Russia}
\affiliation[6]{Ioffe Institute, Polytekhnicheskaya, 26, St. Petersburg, 194021, Russia}
\affiliation[7]{LIRA, Observatoire de Paris-PSL, CNRS, Sorbonne Université, Université Paris Cité, Université Cergy, 92190  Meudon, France}

\abstract{
Solar radio bursts exhibit intricate variability in time, space, and frequency, 
often displaying a rich variety of fine frequency-time structures such as spikes, 
drift pairs, striae in Type III bursts, and herringbone patterns in Type II bursts, etc.
Historically, limited spatial, spectral, and temporal resolution has hindered detailed 
investigation of these narrow‑band, rapidly evolving features, restricting progress 
in identifying their physical origins and underlying processes at these scales. 
Advances in high-time-frequency-resolution solar imaging now offer transformative 
opportunities. Recent sub-second imaging spectroscopy has revealed that many fine structures 
challenge existing theoretical models, pointing to the need for new frameworks 
and a reassessment of current interpretations. 
The Square Kilometre Array (SKA), 
with its full‑Stokes imaging spectroscopy at sub‑second cadences, 
will provide unprecedented data essential for resolving these long-standing questions.
These capabilities promise to significantly deepen our understanding 
of electron acceleration and transport, magnetic reconnection, and coronal plasma turbulence, 
thereby advancing our knowledge of solar energetic processes 
and improving assessments of their space‑weather impacts.
}
\begin{document}

\maketitle
\setcounter{tocdepth}{1}
{\small \tableofcontents}

\section{Introduction}
\begin{figure}[htb!]
    \centering
    \includegraphics[width=0.9\linewidth]{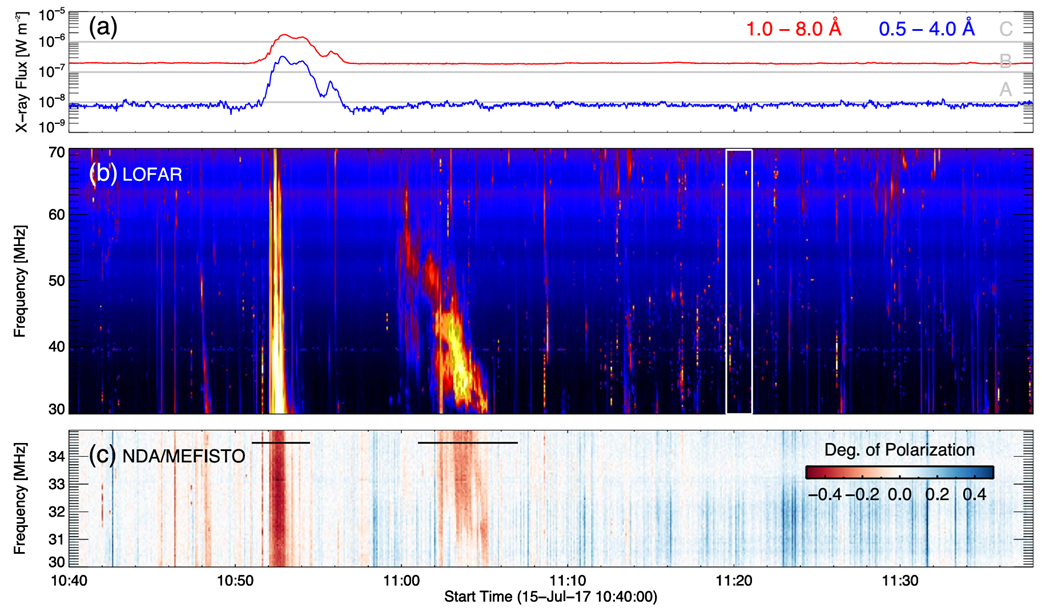}
    \caption{Solar flare with X-ray emission and type III and type II
    radio emissions. (a) GOES X-ray light curves. (b) LOFAR dynamic
    spectrum showing a series of bright Type III bursts followed by
    type II bursts \citep{2020ApJ...893..115C} and myriad of fine
    structures.  
    (c) Dynamic spectrum of the circular polarization from the NDA.
    The image is adapted from \citep{2021ApJ...917L..32C}.}
    \label{fig:overview}
\end{figure}

\begin{figure}[htb!]
    \centering
    \includegraphics[width=0.9\linewidth]{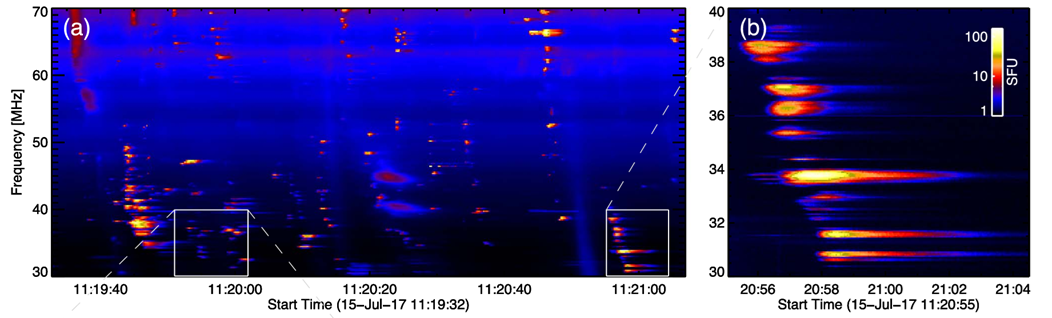}
    \caption{Zoomed in dynamic spectra. (a) Numerous fine structures
    are visible in the region bounded by white white box in Figure
    \ref{fig:overview} (b) Zoomed-in Type IIIb burst near 11:21 UT.
    The image is adapted from \citep{2021ApJ...917L..32C}.}
    \label{fig:zoom_in_overview}
\end{figure}

During periods of sporadic solar activity, the Sun efficiently
converts magnetic energy into the kinetic energy of particles. 
While similar processes are observed throughout the Universe 
(e.g., stellar coronae, 
in planetary magnetospheres and ionospheres, 
or in accretion disks), 
solar phenomena are unique due to their proximity, 
which allows for unprecedentedly detailed studies
\citep[e.g.][as reviews]{2011SSRv..159....3D,2011SSRv..159..107H,2011SSRv..159..301K,2017LRSP...14....2B}.
Signatures of energized thermal and non-thermal electrons are observed
in hard X-rays, ultraviolet, and radio domains. Among these, radio
emission offers unmatched temporal resolution, which is essential for
studying solar transient phenomena (see Figure \ref{fig:overview} and
zoomed in Figure~\ref{fig:zoom_in_overview}, 
providing a unique high-time-resolution perspective
on key questions in solar physics. 

One of the most fascinating aspects of solar radio emission is its
complexity and variability. Full-Sun spectroscopic observations have
catalogued a variety of sub-second, relatively narrow-band features
\citep{2011ASSL..375.....C}. 
Notable examples of these fine structures include
spikes, fine structures within type I bursts, type IIIb (or stria)
bursts, pulsations, and drifting pairs. These features have been
studied primarily through spectroscopy, and more recently with
imaging-spectroscopy below 50 MHz. However, while dynamic spectra have
provided detailed information on fine structures, imaging studies
remain scarce or even absent. Notably, imaging observations have
sometimes produced unexpected results. For example, early imaging of
spikes above 400 MHz (though unresolved in time) by 
\citet{2011ASSL..375.....C} suggested that decimetric spikes 
do not originate from coronal X-ray flare sources, 
contrary to previous expectations. 
The first frequency and time resolved observations of
spikes \citep{2021ApJ...917L..32C,2025ApJ...978...73C} show that the
radio source centroid velocities are often superluminal.

The most fascinating features of solar radio emission are its
complexity and variability. Using full-Sun spectroscopic observations,
the presence of various sub-second and relatively narrow band features
has been well catalogued \citep{2011ASSL..375.....C}. Bright examples
of fine structures are spikes, fine structure of type I bursts, Type
IIIb (or stria) bursts, pulsations, drifting pairs, etc. These
features have been actively studied mostly spectroscopically, 
and more recently using imaging-spectroscopy below 50~MHz. 
While the fine structures are well studied using the dynamic spectra, 
the imaging studies are scarce or non-existent. 
Importantly, imaging observations
often show rather puzzling results. Early imaging observations of
spikes above 400~MHz but unresolved in time
\citet{2009A&A...499L..33B} suggest that decimetric spikes do not
originate from coronal X-ray flare sources contrary to previous
expectations. The first frequency and time resolved observations of
spikes \citep{2021ApJ...917L..32C,2025ApJ...978...73C} show that the
radio source centroid velocities are often superluminal.

The main challenge in frequency-time resolved observations of fine
structures arises from the high variability of the emission (see
Figure \ref{fig:overview}), which necessitates imaging 
at many frequencies simultaneously and with high time resolution. 
Such observations of fine structures have only recently been performed
\citep{2017NatCo...8.1515K} with the Low Frequency Array 
\citep[LOFAR,][]{2013A&A...556A...2V}. 
As a result, time-resolved imaging of fine structures 
has long been a significant challenge.

The Square Kilometre Array 
\citep{braun2019anticipatedperformancesquarekilometre} 
presents new opportunities for discoveries in solar physics 
\citep{2015aska.confE.169N,2019AdSpR..63.1404N}, 
and in particular for the study of solar radio burst fine structures. 
This chapter summarizes recent observations and challenges, 
highlighting the unique opportunities
that the SKA will provide.

\section{Spike radio bursts}

Narrowband, short-duration fine structures known as radio spike bursts
are often observed independently of broadband emission, making them a
notable phenomenon. They appear across a wide frequency range, from
gigahertz \citep{1986SoPh..104..117S, 1992A&AS...93..539B,
1994A&A...285.1038K, 2011ApJ...743..145C,
2011SoPh..273..377D,2015Sci...350.1238C,2021ApJ...911....4L} and
hundreds of megahertz \citep{1972A&A....16...21T, 1990A&A...231..202G,
1995A&A...302..551K, 2016A&A...586A..29B}, down to tens of megahertz
\citep{1994A&A...286..597B, 2014SoPh..289.1701M, 2016SoPh..291..211S,
2023ApJ...946...33C}. Their high brightness temperatures link them to
coherent emission mechanisms such as electron cyclotron maser emission
\citep{1982ApJ...259..844M, 1992A&AS...93..539B} and plasma emission
\citep{1972A&A....17..267T, 2014SoPh..289.1701M,
2021ApJ...917L..32C,2024OJAp....7E..51M}.

The prevalence of spikes, especially at and above decimetre
wavelengths, can cause radio interference at Earth, 
and be particularly problematic for GPS/GLONASS 
noise \citep{2008SpWea...610D07C,
2009RaSc...44.0A08K, 2011ApJ...743..145C}. The spikes occur in several
solar activity environment, i.e. noise storms, 
Type III storm, Type II or Type IV solar radio bursts. 
At a few hundred MHz frequencies, the spikes could
occur with type III bursts at frequencies above the type III starting
frequency and can coincide with HXR emission
\citep{1991A&A...251..285G}. This has led to the view that spikes are
a direct signature of the acceleration of non-thermal electrons
\citep{1982A&A...109..305B}. Their sporadic appearance in dynamic
spectra can form clusters of thousands of spikes, which may suggest
that the acceleration region is fragmented and that energy release
could occur across many small sites \citep{1985SoPh...96..357B}.
At decametre frequencies, recent imaging results
show that spike bursts share spatial and morphological features with
type IIIb striae. This suggests that decametre spikes arise from
propagating electron beams and plasma emission, modulated by density
fluctuations in the intervening plasma, producing radio emission
across narrow spatial regions \citep{2021ApJ...917L..32C,
2023ApJ...946...33C}.

\subsection{Time-frequency characteristics}

\begin{figure}[htb!]
    \centering
    \includegraphics[width=0.49\linewidth]{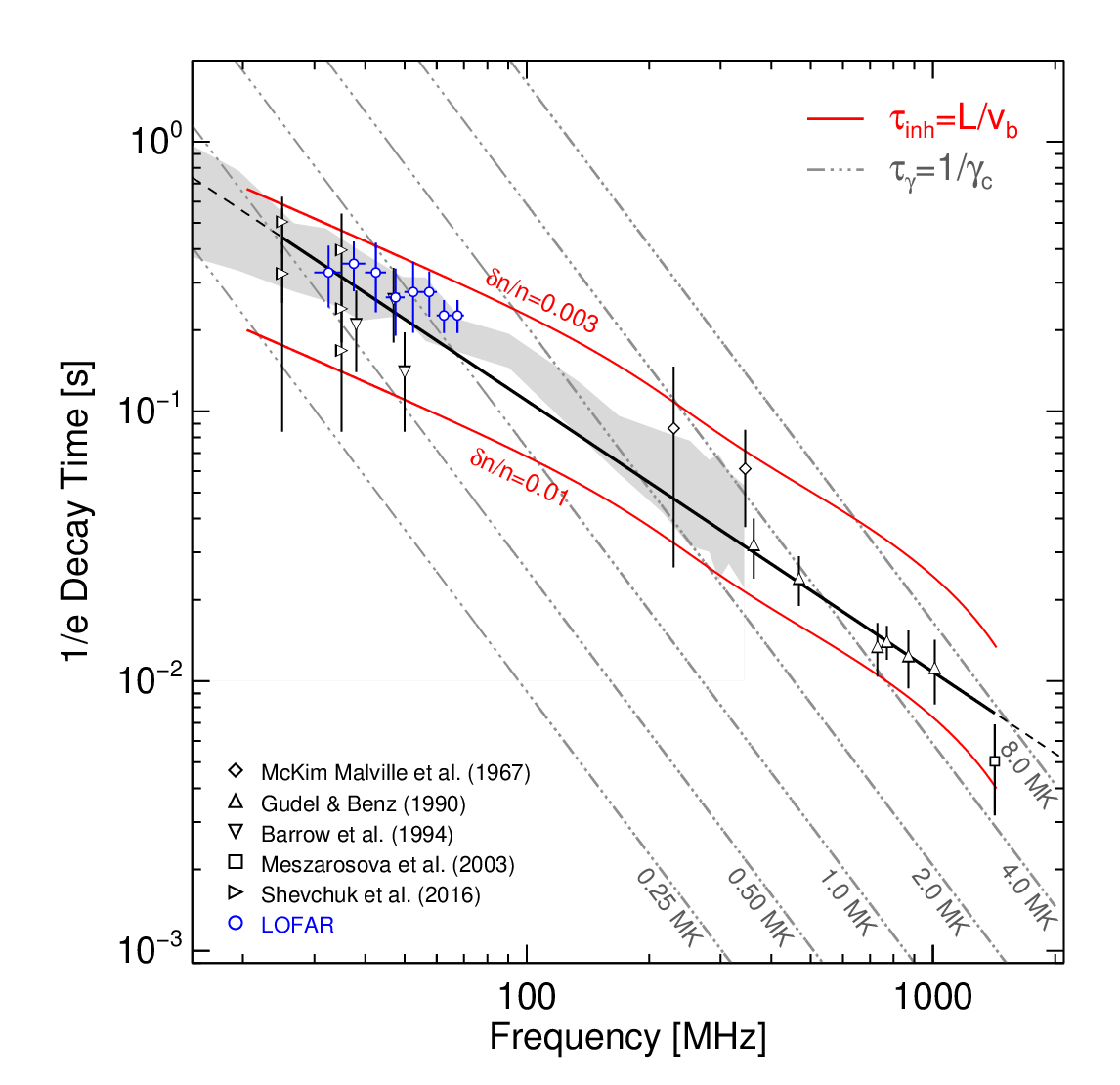}
    \includegraphics[width=0.46\linewidth]{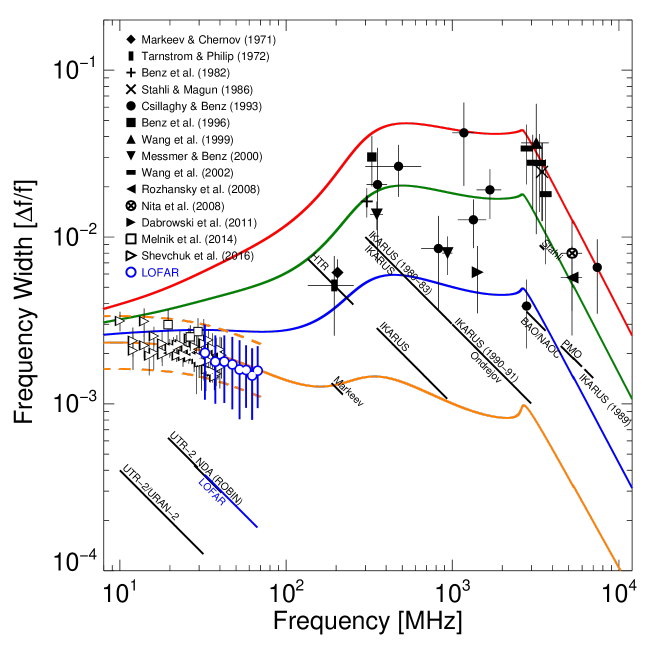}
    \caption{\textit{Left:} Average spike 1/e decay times against
    frequency. The solid black line represents the power-law fit to
    the data given by $\tau_d=(11.22\pm1.9)f^{-1.01\pm0.03}$. The
    dash-dotted gray lines represent the plasma collision time for
    various coronal temperatures. The gray region shows the typical
    scattering decay time where the spread accounts for halving and
    doubling of the scattering rate \cite[see][]{2023ApJ...956..112K}.
    The red curves show the inhomogeneity time. \textit{Right:}
    Average spike bandwidth ratio $\Delta{f}/f$ where $\Delta{f}$ is
    given at the FWHM level. The solid black lines show the instrument
    spectral resolution. The coloured lines denote the bandwidth
    derived from the Langmuir-wave dispersion relation for various
    magnetic field profiles. Figures adapted from
    \cite{2023ApJ...946...33C}.}
    \label{fig:spike_decay_bandwidth}
\end{figure}

Spikes are likely the shortest duration radio emissions observed from
the Sun at decametre wavelengths. At decimetric frequencies, the
timescales are on the order of tens of milliseconds, increasing to
hundreds of milliseconds to $\sim 1$~s at decametre frequencies.
Earlier studies linked the brief lifetime of spikes to collisional
damping in the plasma \citep[e.g.][]{1990A&A...231..202G}. However,
recent observations reveal significant radio-wave scattering acting on
these sub-second bursts. The decay time of spikes follows a $1/f$
relation across a wide frequency range (Figure
\ref{fig:spike_decay_bandwidth}), consistent with radio-wave
scattering theory \citep{2023ApJ...956..112K}. Near 30~MHz, scattering
appears to dominate the observed profiles, suggesting that the
intrinsic emission timescale is much shorter on the order of tens of
milliseconds rather than $\sim1$~s as observed
\citep{2021ApJ...917L..32C}. This suggests that the intrinsic
timescale of the emission may be shorter than what is inferred from
observations. The SKA's planned temporal imaging resolution 
of $1$~ms will greatly enhance the ability 
to resolve high frequency spikes over the broad range of frequencies 
(Figure~\ref{fig:spike_decay_bandwidth}). 
This capability will enable detailed studies of emission 
and acceleration timescales in flaring
regions, and may also reveal the fine structures 
that contribute to type I noise storm continua.

Spike bandwidths are typically in the range $\Delta{f}/f\simeq10^{-3}
- 10^{-2}$, hence often requiring high frequency resolution. 
They usually have symmetrical profiles in frequency, with bandwidth
increasing at higher frequencies. Above $\sim200$~MHz, spike bandwidth
ratios are an order of magnitude larger 
than those below $100$~MHz. 
This broadening can be attributed to the Langmuir wave
dispersion relation in a magnetized plasma
\citep[e.g.][]{2016SoPh..291..211S, 2023ApJ...946...33C}, which
depends on the local magnetic field strength. However, a major
limiting factor in past bandwidth measurements at decimetric
frequencies has been the spectral resolution of the observing
instruments, which may have led to overestimates in statistical
averages (Figure \ref{fig:spike_decay_bandwidth}). 
For spikes produced
by plasma emission and modulated by density inhomogeneities, the
bandwidth can be used to estimate the size of the emitting region
(that is, the size of the density inhomogeneity) as $\Delta{r}\simeq
2L(\Delta{f}/f)$ where $L$ is a characteristic density scale height
\cite[e.g.][]{2017NatCo...8.1515K}. SKA-Low has a standard spectral
resolution of $5.4$~kHz with a maximum resolution of 226~Hz in zoom
mode \citep{2022JATIS...8a1018H}.
In combination with imaging, this improvement will allow radio
fine structures such as spikes, 
repeating bursts \citep{2026NatCo..17.5131M}, or noise storms \citep{2026arXiv260531450C}
to probe smaller regions of plasma turbulence 
and help to constrain emission region properties and sizes.

\begin{figure}[htb!]
    \centering
    \includegraphics[width=0.35\linewidth]{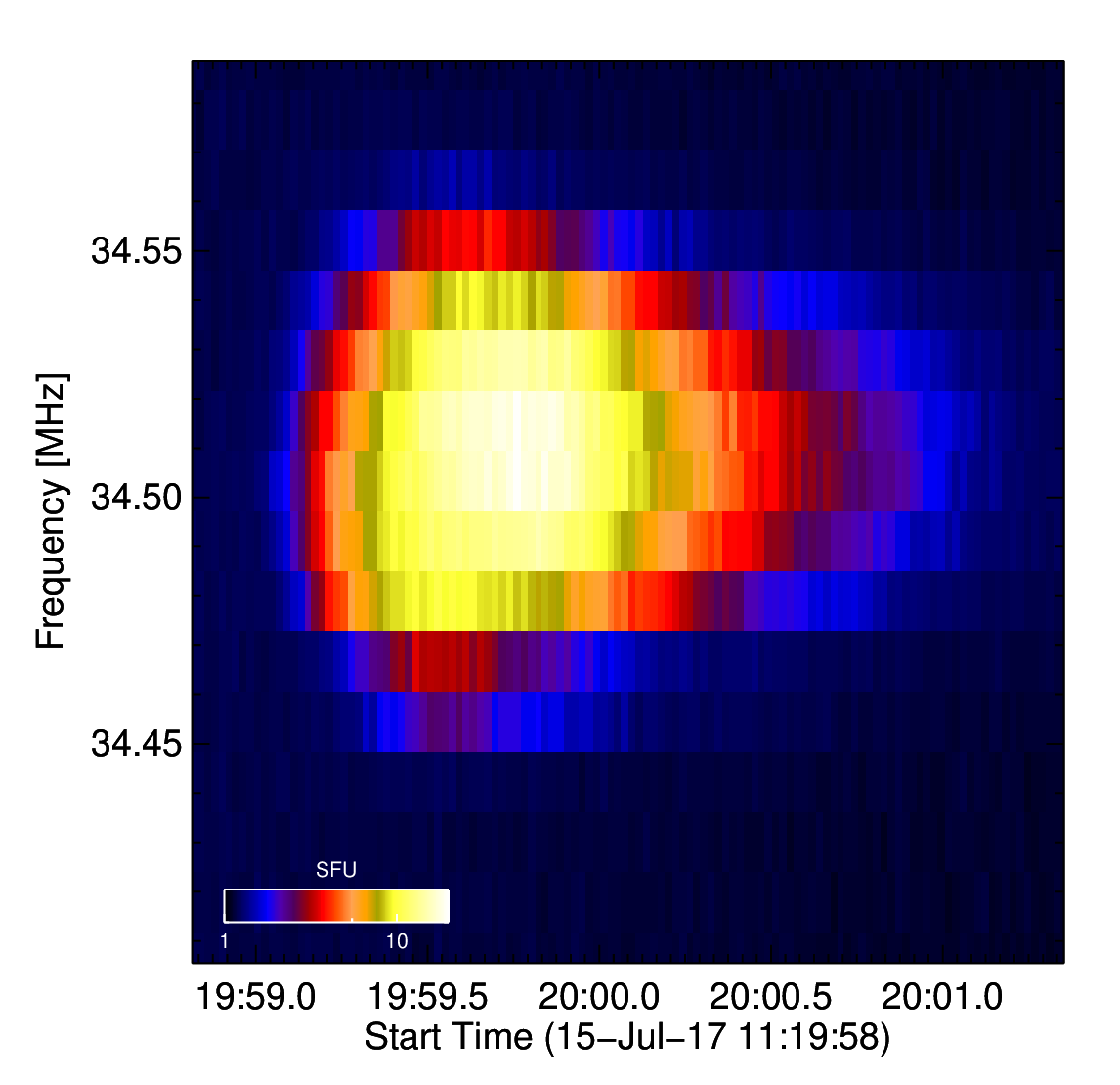}
    \includegraphics[width=0.35\linewidth]{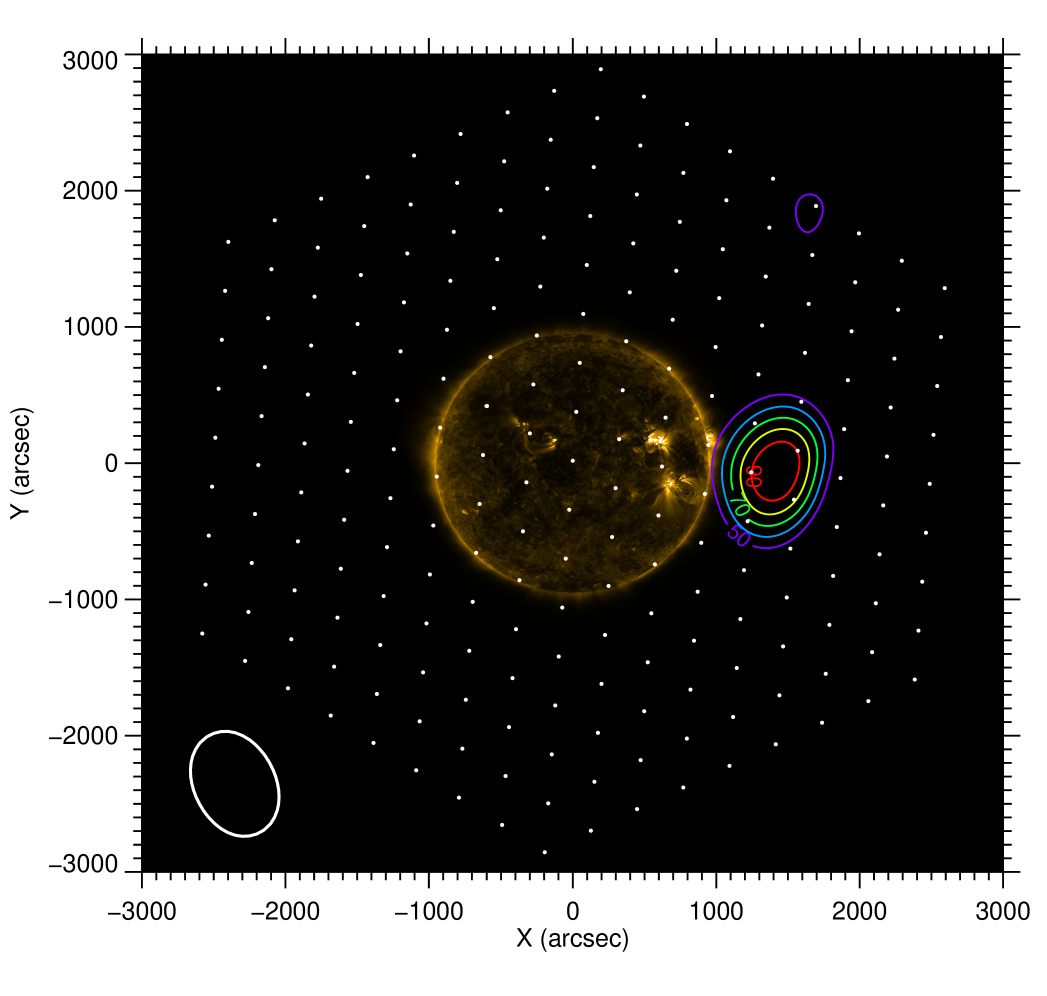}\\
    \includegraphics[width=0.35\linewidth]{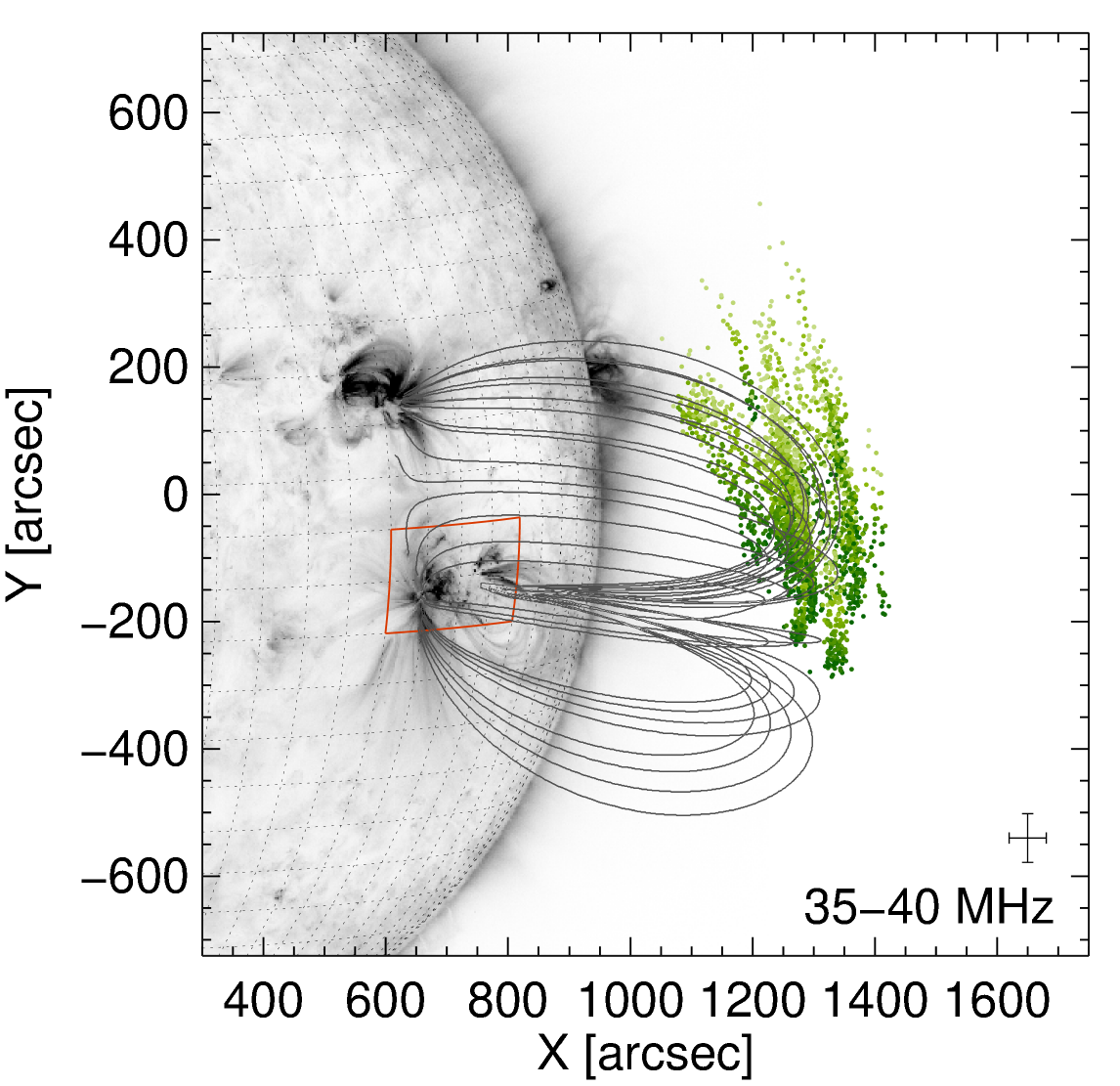}
    \includegraphics[width=0.35\linewidth]{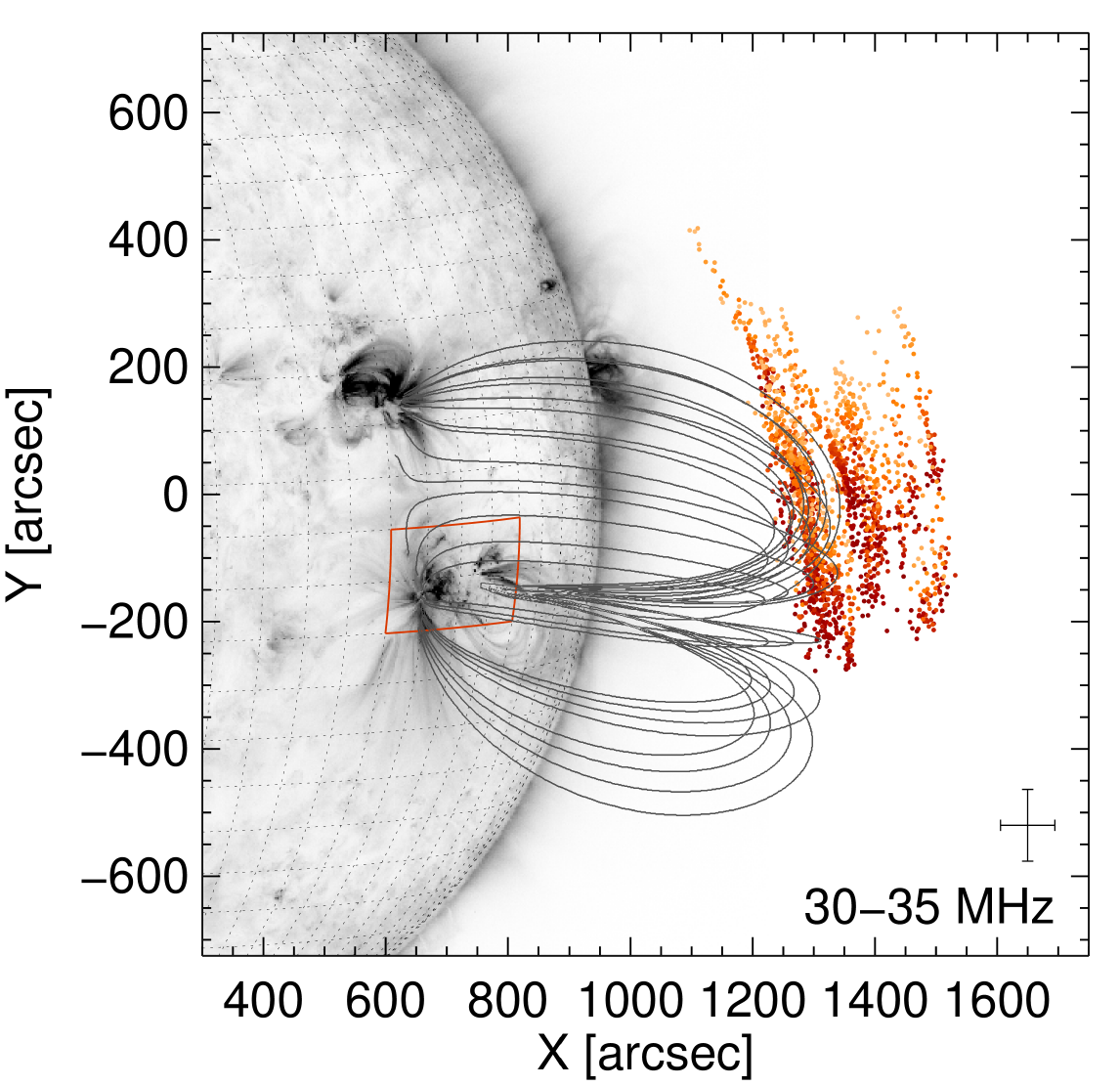}
    \caption{\textit{Top Left:} LOFAR dynamic spectrum of an
    individual spike. \textit{Top Right:} LOFAR image of the spike
    peak flux at 34.5~MHz. The oval in the lower left corner
    represents the beam size. \textit{Lower Left:} LOFAR observed
    centroid positions of hundreds of spikes between 35-40 MHz with
    time increasing from dark to light. \textit{Lower Right:} As in
    the lower left panel but for spikes observed between 30-35~MHz.
    Lower panels adapted from \cite{2023ApJ...946...33C}.}
    \label{fig:spike_imaging}
\end{figure}

Above a few hundred megahertz, spike frequency drift rates range
between $10^2-10^4$~MHz~s$^{-1}$ and show a similar frequency
dependence to type III bursts. Near 30 MHz, however, spike emission is
strongly affected by temporal broadening due to radio-wave scattering,
which smears the signal in dynamic spectra and reduces the apparent
drift-rate \citep{2025ApJ...978...73C}. Even with the spectral
resolution of LOFAR, decametre spikes tend to cover a modest number of
frequency channels (Figure \ref{fig:spike_imaging}), making it
difficult to track the drift of the intensity peak across the
channels. Using the same approach used type IIIb striae
\cite[e.g.][]{2018SoPh..293..115S}, \cite{2023ApJ...946...33C} measure
absolute spike drift rates from near zero to a few tens of kilohertz
per second. The dilution of the drift rate affects estimates of the
plasma temperature \citep{2025ApJ...978...73C}, a key parameter
influencing the drift of fine structures produced through the plasma
emission process \citep{2021NatAs...5..796R}. The enhanced spectral
resolution of SKA-Low will enable more precise tracking of spike peak
flux across frequency channels, improving estimates of coronal
temperatures and the role of scattering in shaping the observed burst
dynamics.

\subsection{Imaging}

Radio imaging observations of spikes remain limited because few
instruments can provide the necessary combination of spatial,
spectral, and temporal resolution. The first resolved imaging of
spikes was achieved with LOFAR near 30 MHz \citep{2021ApJ...917L..32C,
2023ApJ...946...33C}, showing that spike source sizes follow
radio-wave scattering predictions consistent with strong anisotropy in
density turbulence \citep{2023ApJ...956..112K,2025ApJ...991L..57K}. At a fixed frequency
over time, the spike bursts displayed substantial positional 
shift across the sky-plane --- up to nearly a solar radius 
in less than $1$~s along the direction of the guiding magnetic 
field. This demonstrates the dominant role of radio-wave scattering, which can produce
superluminal apparent source motion. The LOFAR observations, 
performed in beam-formed mode \citep{2022ApJ...925..140G}, 
enabled tracking of spike sources on tens of
millisecond timescales but with a modest angular resolution of
$\sim9$~arcmin at 30 MHz. As a result, any substructure within
individual bursts or multiple sources could not be resolved. 

At 50~MHz, SKA-Low configured with a $\sim6$~km baseline would be able
to tile a region of  $2.5$~R$_\odot$ radius centred on the Sun using
500 beams, achieving an angular beam width of $3.5$~arcmin---nearly a
factor of two improvement over LOFAR's $5.9$~arcmin using 217 beams
with a $3.5$~km baseline. Moreover, convolution of the instrumental
beam affects the apparent size and shape of radio sources, and
improved angular resolution will improve this distortion. Since source
broadening occurs mainly perpendicular to the magnetic field due to
anisotropic scattering \citep{2019ApJ...884..122K}, SKA-Low's enhanced
imaging may reveal source orientations that can be used to infer the
direction of the magnetic field.

\section{Drift-pair bursts}
Drift-pair bursts are a rare type of fine structure that are observed
sometimes in low-frequency solar radio emission (in the frequency
range of about 10-100 MHz). They were first identified as a separate
class of fine structures by \citet{1958AuJPh..11..215R}. The bursts
appear in the dynamic spectra as two parallel frequency-drifting
narrowband stripes separated in time, with a typical delay between the
components of about 1-2 s. The frequency drifts can be both positive
(which are more common) and negative. The drift rates tend to increase
with the emission frequency, being generally of about 1-2 MHz s$^{-1}$
at the frequency of $\sim30$~MHz. The emission intensities are up to
hundreds of sfu, and the brightness temperatures exceed $10^{10}$~K,
indicating a coherent emission mechanism. The circular polarization
degree varies from very low to $\sim50\%$. The bursts usually occur
not individually, but form large irregular groups (clusters).

\subsection{Time-frequency characteristics}

The most intriguing feature of drift-pair bursts is their double
structure, where the second component of a pair looks like a
repetition of the first one with the same frequency range and drift
rate, although sometimes with a slightly different intensity. Also,
the higher-frequency component generally has a higher polarization
degree. The duration of each component at a single frequency is of
$\sim1$~s, which corresponds to an instant bandwidth of $\sim1$~MHz.
The delay between the components tends to decrease slightly with
frequency. The bursts with negative frequency drift are usually
shorter and more narrowband than those with positive frequency drift.

\subsection{Imaging}
\citet{1979PASA....3..379S}, using the Culgoora radioheliograph, first discovered that the sources of both components of a drift-pair
burst (at the same frequency) coincide spatially.  
Later, \citet{2019A&A...631L...7K}, 
using multi-frequency imaging observations with LOFAR, have found that the sources of both
components of a drifting pair propagate 
(with a certain delay) in the
same direction along the same trajectory. At a fixed frequency, at the
decay phases of both components, the emission source demonstrates an
outward radial motion (with the speed of about $c/3$) and an increase
in size; this behaviour is reminiscent of the source dynamics in type
IIIb bursts \citep{2017NatCo...8.1515K}.

Since the first observations, the formation of drift-pair bursts was
attributed to the radio echo effect, when the second component
represents a signal reflected from lower layers of the solar corona.
On the other hand, a direct (regular) reflection in a stratified
corona cannot reproduce the observed features, because it would
provide different apparent source positions and different intensities
and time profiles of the direct and reflected signals. Using numerical
simulations, \citet{2020ApJ...898...94K} have recently demonstrated
that the key factor in the formation of drift-pair bursts could be
scattering of the emission on anisotropic plasma density fluctuations.
The combination of reflection and anisotropic scattering (turbulent
radio echo) has been shown to provide similar source sizes and locations and
comparable intensities of the direct and reflected burst components,
as well as to reproduce the source dynamics and the delays
($\sim1-2$~s) between the components. This model also explains why
drift-pair bursts are only observed below $\sim100$~MHz: at higher
frequencies, the reflected component becomes damped due to collisional
absorption. Thus anticipated high-quality observations of drift-pair
bursts with SKA offer a fascinating opportunity to diagnose the plasma
turbulence in the outer solar corona.
\begin{figure}[htb!]
    \centering
    \includegraphics[width=0.9\linewidth]{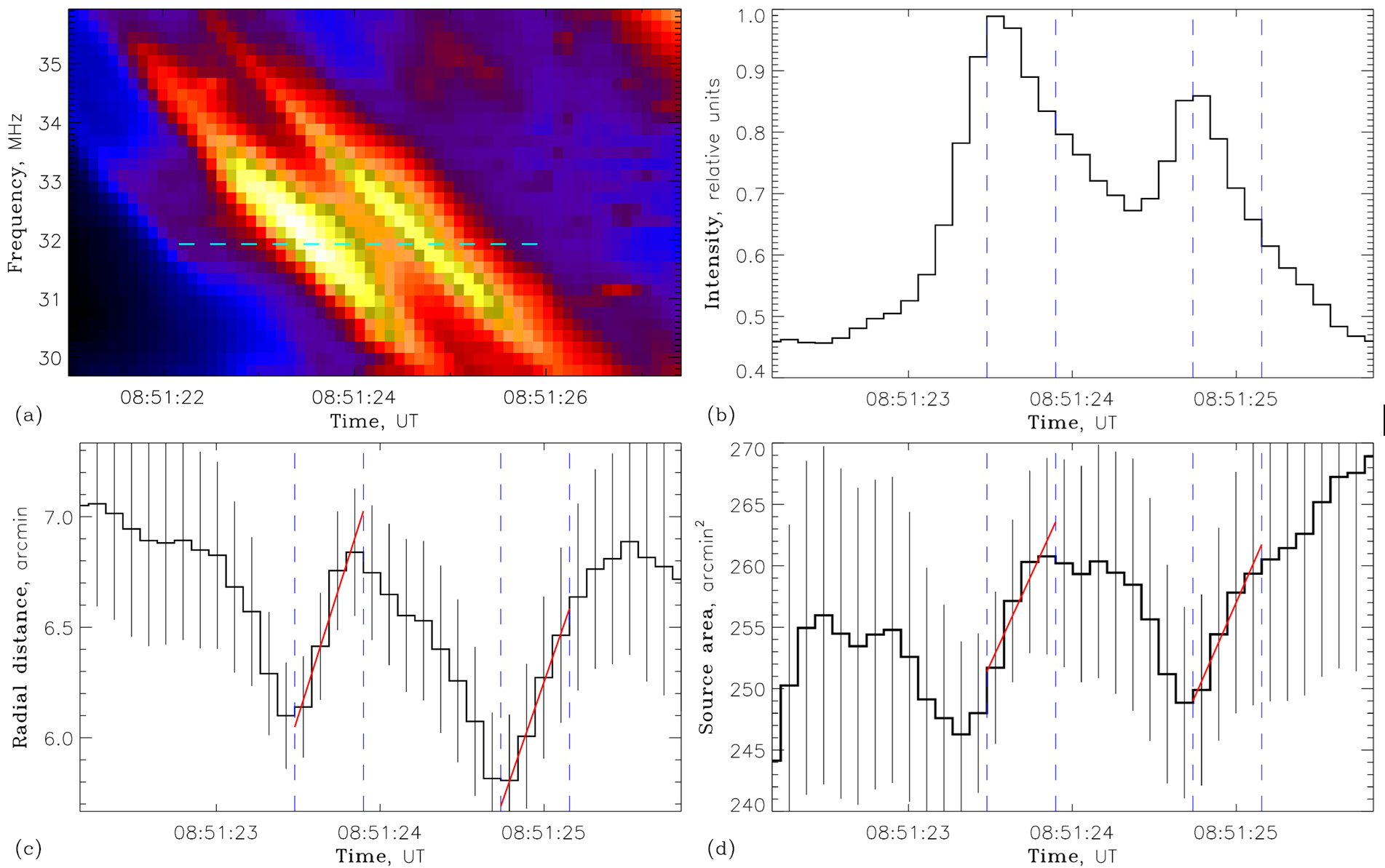}
    \caption{(a) Dynamic spectrum of solar radio emission with a
    drift-pair burst recorded with LOFAR on 2017 July 12
    (Sun-integrated, in relative units). (b) Time profile of the radio
    flux at a single frequency (32 MHz). (c) Corresponding time
    profile of the visible radio source position (distance from the
    solar disk center). (d) Corresponding time profile of the visible
    radio source area (at half-maximum level). Red lines represent
    linear fits to the time profiles of the source parameters in the
    intervals shown. The error bars represent one standard deviation.
    Figure adapted from \citep{2019A&A...631L...7K}.}
    \label{fig:drifting_pairs}
\end{figure}

Besides, the origin of the emission itself remains unidentified yet.
The emission is likely produced by non-thermal electrons 
as suggested by the high brightness temperatures; 
Interestingly, \citet{2019A&A...631L...7K} detected drift-pair bursts 
in absence of any flaring activity, while the emission source 
was located at a boundary of a coronal hole. 
Thus observations of drift-pair bursts can
shed light on the magnetic reconnection and energy release processes
in the otherwise quiet outer solar coronal regions, where solar wind
acceleration begins. The frequency drift rates of drift-pair bursts
are consistent with propagating MHD waves or whistler 
packets \citep[see discussion][]{2020ApJ...898...94K}, 
and hence observations of the spatial, spectral, 
and temporal evolution of these bursts can be used 
to diagnose the magnetic field and plasma structures in their sources.

\section{Type III striae}
Type III bursts are a form of transient radio emission that quickly 
drifts from frequencies as high as hundreds of MHz or GHz down 
to as low as tens of kHz. 
The drift rate that scales approximately as 
$df/dt = 0.01f^{1.84}$ \citep{1973SoPh...29..197A}, 
is caused by energetic electrons propagating out from the Sun 
through plasma with the decreasing background plasma density 
of the solar corona and solar wind. Type III stria
\citep{1978A&A....70..685A,1979SoPh...62..145A,2010AIPC.1206..445M},
also known as type IIIb bursts 
\citep{1972A&A....20...55D,2018SoPh..293...26M,2017NatCo...8.1515K}, 
are fine structures that are present
within the bulk frequency drift of type III bursts. Stria are stripes
in the dynamic spectra that drift slower than the type III burst
($df/dt \approx 0.1$ MHz s$^{-1}$ at 30 MHz) with short durations
(about 1~s at 30 MHz) and have a characteristic frequency bandwidth
$\Delta f/f \approx 10^{-3}$, independent of frequency
\citep{2018SoPh..293..115S}.

Stria are typically observed at frequencies above 10 MHz due to the
higher resolution of ground-based telescopes, 
but have recently been observed below 1~MHz frequencies
\citep{2020ApJS..246...49P,2021ApJ...915L..22C} and even below 100 kHz
\citep{2025ApJ...985L..27K}.  The source size of stria has been
estimated to be 50-400 arcmin$^2$ around 25 and 40 MHz
\citep{2017NatCo...8.1515K,2020ApJ...905...43C,2020A&A...639A.115Z}.
The driver for type III stria is believed to be density turbulence in
the background plasma \citep{1975SoPh...40..421T}. Enhanced
resolutions obtained from LOFAR observations have found that the power
spectral density of type III stria have a -5/3 spectral index for both
fundamental and harmonic emission
\citep{2018ApJ...856...73C,2021NatAs...5..796R}, like the inertial
range of density turbulence measured in the solar wind
\citep{1991RvMA....4..145M,2013LRSP...10....2B}. Power spectral
densities detected in herringbone type II bursts have shown similar
properties \citep{2021ApJ...921....3C,2023ApJ...952...51K}.

A theoretical framework for the generation of striae was provided by
\citep{2021NatAs...5..796R}, supported by numerical simulations. They
demonstrated the fine structure could be caused by modulation 
in the growth rate of beam-driven Langmuir waves that in turn modulates 
the radio emission.
The drift rate of stria is attributed to the motion of intense
clumps of Langmuir waves in a turbulent medium, at the Langmuir wave
group velocity $v_{gr} = 3v_{Te}/v_b$, where $v_{Te}$ is the electron
thermal velocity and $v_b$ is the bulk speed of the electron beam. The
intensity of radio emission can then be used to infer the level of
density fluctuations parallel to the magnetic field via $\Delta I/I =
(v_{Te}/v_b)^2 \Delta n/n$, where $v_{Te}$ is the electron thermal
speed and $v_b$ is the electron beam speed. Values of $\Delta n/n =
3\times 10^{-3}$ were found in the corona from stria, with similar
(somewhat larger) values of $\Delta n/n = 8\times 10^{-3}$ found from
the fine structure in herringbones associated with a type II radio
burst \citep{2021ApJ...921....3C}. Other studies have also approximated the
level of density fluctuations $\delta n/n$ based upon the observed
value of $\Delta f/f $ in stria, finding similar values from $10^{-2}$
to $10^{-3}$
\citep{2017SoPh..292..155M,2018SoPh..293..115S,2025ApJ...985L..27K}.
Anisotropic scattering of radio waves off density fluctuations has
been shown to modify the properties of stria
\citep{2017NatCo...8.1515K,2019ApJ...884..122K,2020ApJ...905...43C}.
The scattering can cause a non-radial fixed-frequency source motion
over time \citep{2023ApJ...946...33C}.  Additionally, the apparent
drift rate of stria can be reduced \citep{2025ApJ...978...73C},
depending on the magnetic field direction and level of anisotropy of
the density fluctuations.

\section{Type II bursts fine structures}
\begin{figure}[htb!]
    \centering
	\includegraphics[width=0.95\columnwidth]{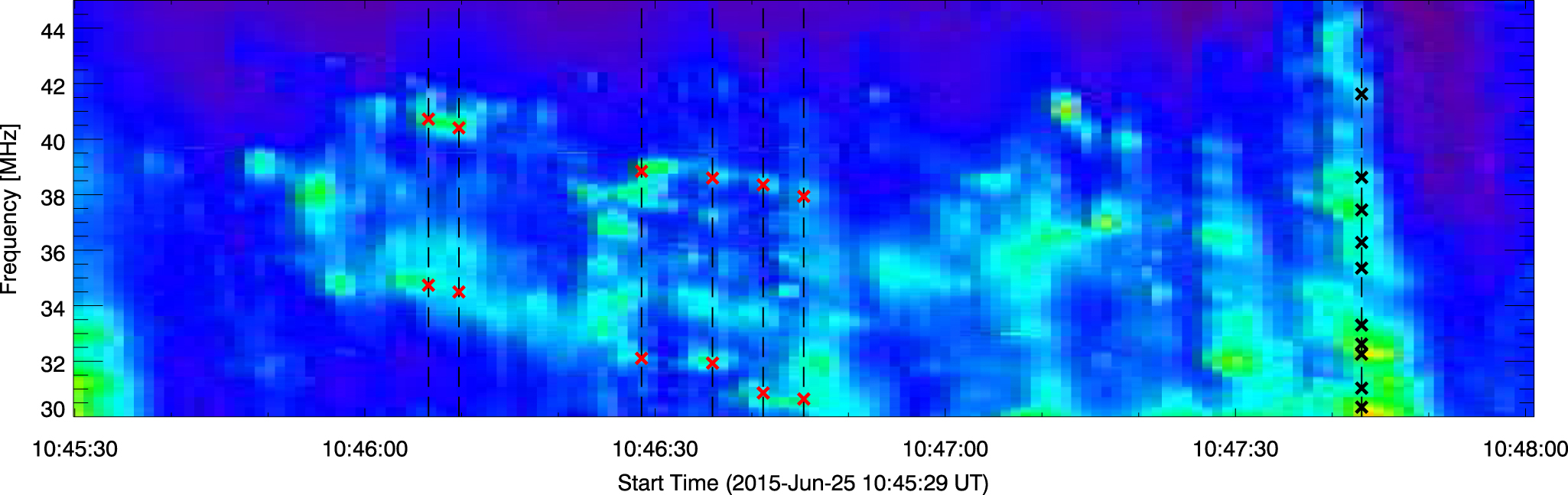}
    \caption{ Type II solar radio burst exhibiting band splitting and
    highly-fragmented bands. LOFAR imaging spectroscopy observations
    allowed for the simultaneous imaging of both subbands at several
    moments in time, as shown by the red crosses. Figure taken from
    \cite{2018ApJ...868...79C} and reproduced by permission of the
    AAS. }
    \label{fig:typeII_2018}
\end{figure}

One of the most characteristic solar radio emissions are the so-called
Type II radio bursts, displaying a much slower frequency-drift rate
than other radio signatures \citep[$\lesssim
-1$~MHz~s$^{-1}$;][]{1950AuSRA...3..399W}. 
At decametric and metric wavelengths ($\gtrsim$10~MHz) they tend 
to have a life span of several
minutes and relatively short instantaneous bandwidths. 
Type II bursts have been associated with shock waves 
traversing the heliosphere,
normally driven by Coronal Mass Ejections \citep[CMEs;
e.g.][]{1999SoPh..187...89C}. They also exhibit larger-scale
structures, particularly band splitting---the separation of the Type
II burst backbone into thinner, parallel lanes---and
herringbones---Type-III-like bursts that appear to emanate from the
backbone with opposing frequency-drift rates on each side
\citep[e.g.][]{1959AuJPh..12..327R}. However, fine structures within
Type II bursts have also been identified since the very early
observations, with \cite{1959AuJPh..12..327R} noting that Type II
bursts ``are rarely smooth and continuous but fluctuate in intensity
over periods of seconds'', giving the appearance of a fragmented or
patchy backbone (see Figure~\ref{fig:typeII_2018}).

The mechanisms generating the larger-scale Type II structures (band
splitting and herringbones) have been a source of long-standing debate
\citep[e.g.][]{1974IAUS...57..389S, 1975ApL....16R..23S,
1983ApJ...267..837H, 2005A&A...441..319M}. There is no consensus on
the mechanisms leading to the acceleration of energetic electrons and
subsequent generation of radio emissions along shocks. Detailed
studies of the finer and fainter structures observed within these Type
II emissions could shed light onto the generation mechanisms
responsible for the various morphologies. Thus, the ability to resolve
and study these fine structures in detail (both spectroscopically and
in images) is key to addressing these long-standing debates. The
sensitivity of the SKA, along with the high temporal and spectral
resolutions, and its ability to conduct imaging spectroscopy
observations, will play an integral role in such studies. Imaging
spectroscopy is particularly beneficial as it allows us to identify
the locations of the Type II radio emissions corresponding to each
spectral measurement and explore the shock conditions and mechanisms
that lead to the acceleration of electrons. For example, a major
distinction in the two mainstream band-splitting models is whether the
radio sources of the two subbands are co-spatial or physically
separated along the shock \citep{1974IAUS...57..389S,
1975ApL....16R..23S, 1983ApJ...267..837H}. Due to instrumental
limitations inducing time-delay ambiguities in the observations, it
has been challenging to distinguish between the two scenarios.
\cite{2018ApJ...868...79C} used LOFAR's imaging spectroscopy
observations to conduct the first simultaneous imaging of both
subbands of a split-band Type II burst (Figure~\ref{fig:typeII_2018}),
eliminating these ambiguities that prevented the simultaneous
identification of the subband source locations.
\begin{figure}[htb!]
    \centering
	\includegraphics[width=0.8\columnwidth]{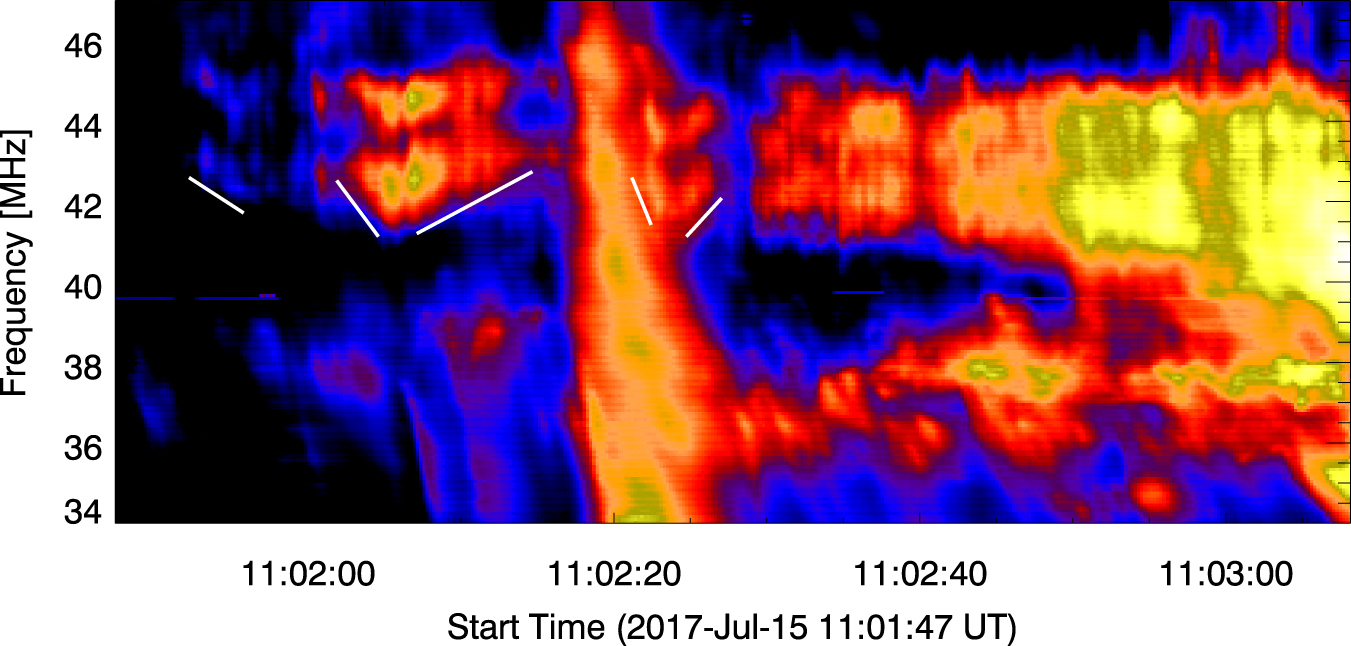}
    \caption{Intriguing fine structures observed within the stationary
    part of a Type II burst that transitions from a stationary to
    drifting state. The stationary parts of the Type II burst consist
    of two bands, both of which exhibit band splitting. The white-line
    annotations highlight fine structures that alternate between
    negative and positive frequency-drift rates. Figure taken from
    \cite{2020ApJ...893..115C} and reproduced under
    \href{https://creativecommons.org/licenses/by/4.0/}{CC BY 4.0}. }
    \label{fig:typeII_2020}
\end{figure}

Observations and identifications of various fine structures embedded
within Type II bursts have increased with the commissioning of
sensitive telescopes with very high temporal and spectral resolutions.
\cite{2018A&A...609A..41M} used high-resolution observations 
by URAN-2 \citep{2005KFNTS...5...43B}
to examine the ``head'' and ``tail'' sub-structures of herringbones
\citep{1987SoPh..111..365C}. \cite{2019A&A...624A..76A} studied a
large number of the spike-like fine structures observed within Type II
bursts, recorded between 270--450~MHz by the ARTEMIS-IV 
\citep{2006ExA....21...41K} spectrograph.
They reported an average duration of 96~ms (with a standard deviation
$\sigma = 54$~ms) and an average relative bandwidth of 1.7\% (with
$\sigma = 0.5$\%). An intriguing behaviour of fine structures within
the stationary part of a Type II burst that also exhibited band
splitting was reported by \cite{2020ApJ...893..115C}. 
This Type II burst---recorded with LOFAR and shown in
Figure~\ref{fig:typeII_2020}---transitioned from a stationary to
drifting state, and displayed fine structures that alternate between
positive and negative frequency-drift rates within the stationary
part. Thanks to imaging spectroscopy observations, the Type II radio
sources at various frequencies were imaged at sub-second intervals,
revealing their spatial evolution, which aided in the identification
of a generation mechanism. Overall, the SKA observations will
facilitate the complete examination of Type II solar radio burst fine
structures---which remain largely unexplored---advancing our ability
to probe the shock-acceleration mechanisms and the impacts of
propagation through the turbulent heliospheric medium.

\begin{figure}[htb!]
    \centering
	\includegraphics[width=0.8\columnwidth]{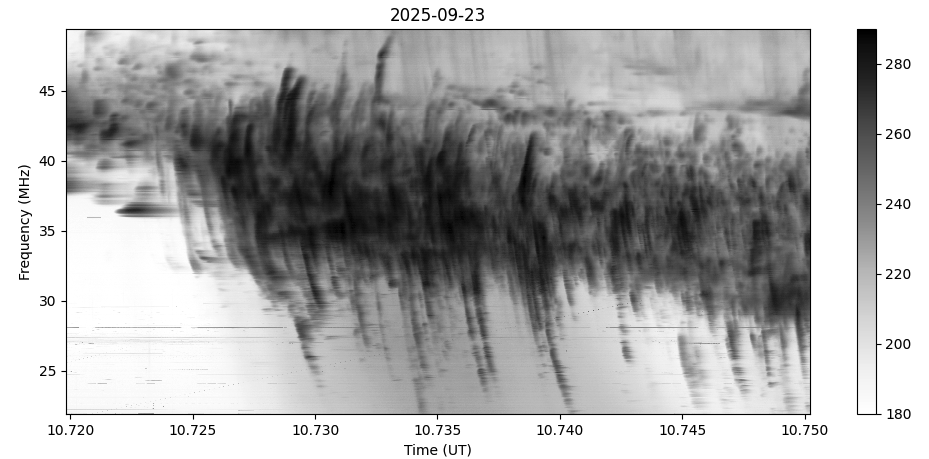}
    \includegraphics[width=0.8\columnwidth]{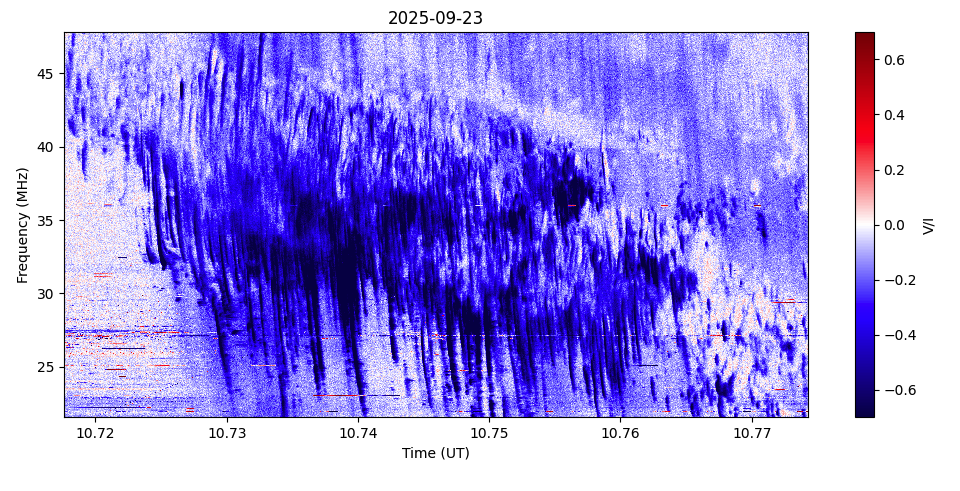}
    \caption{Type II solar radio burst exhibiting herringbone
    structures. \textit{Top:} Stokes I dynamic spectrum \textit{Top:}
    Polarization (Stokes V/I) dynamic spectrum. The spectra are from
    \href{NenuFAR}{https://nenufar.obs-nancay.fr}.}
    \label{fig:typeII_herringbone}
\end{figure}
Herringbone structures in type II solar radio bursts are often
interpreted as direct signatures of electron acceleration at
CME-driven shock fronts
\citep{1966AuJPh..19..209S,1987SoPh..111..365C,2004SoPh..222..151M,2015SoPh..290.2031D}.
In the dynamic spectrum (Figure \ref{fig:typeII_herringbone}), 
these structures appear as numerous short type III-like bursts 
that drift either to higher or lower frequencies, 
resembling fish bones branching off the main 'backbone'.
The detailed properties of herringbones, 
such as their drift rates, durations, imaging locations, 
and polarization characteristics can provide insights 
into the underlying electron acceleration mechanisms. 
\mdseries
\section{Fine structures within flaring regions}

The solar flare observations suggest a link between the X-ray emission
and simultaneous radio emission. However, contrary to previous
expectations decimetric spikes do not originate from coronal X-ray
flare sources \citep{2009A&A...499L..33B}. This vision might be
challenged by the future SKA-mid observations.

\begin{figure}[htb!]
    \centering
    \includegraphics[width=0.99\linewidth]{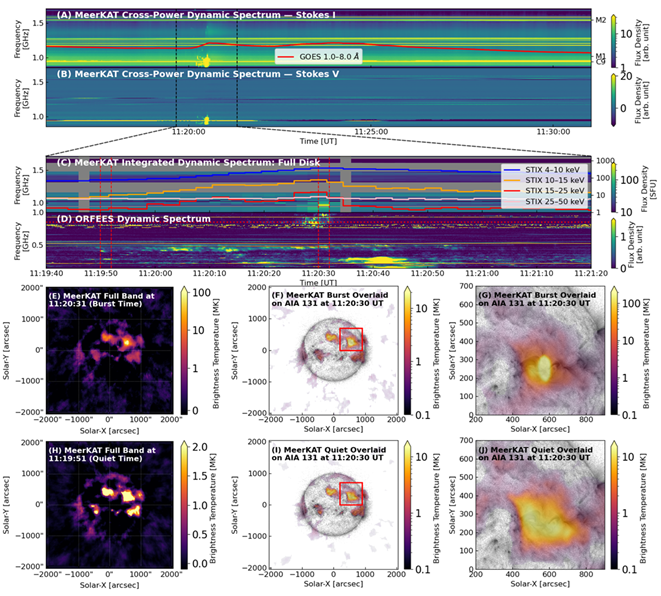}
\caption{ MeerKAT observations of fine structures in a solar flare by
\citet{2026ApJ...998L..46L}. (A) MeerKAT Stokes I cross-power dynamic
spectrum with GOES $1 -8$\,\AA\ soft X-ray flux (red) over-plotted.
(B) MeerKAT Stokes V cross-power spectrum. (C) MeerKAT full-disk
integrated dynamic spectrum during the burst with STIX light curves
overlaid; bad channels and time intervals are flagged. (D) 
ORFEES \citep[][]{2021JSWSC..11...57H}
dynamic spectrum. Vertical dashed black lines in (A) and (B) denote
the interval shown in (C) and (D), while vertical dashed red lines in
(C) and (D) mark the times used for imaging in (E)–(G) and (H)–(J).
The red horizontal line in (D) indicates the low frequency limit of
the MeerKAT band. (E) MeerKAT full-band radio image in brightness
temperature (MK) at the burst peak, a linear colour scale for values
below 1~MK and a logarithmic scale for values above 1 MK is used to
enhance the visualization. (F) Alpha-blended overlay of MeerKAT
emission (inferno colormap, transparency scaled by brightness) 
on an SDO/AIA \citep{2012SoPh..275...17L} 131\,\AA\ image  (grayscale). 
Red boxes outline the regions shown
in the zoomed-in view in (G). (H)–(J) Same as (E)–(G), but for the
pre-burst quiet time.}
    \label{fig:meerkat_fine}
\end{figure}

As the precursor to SKA-Mid, MeerKAT \citep{2016mks..confE...1J} has
already conducted several successful solar flare observations,
demonstrating its capability for stable calibration and high-fidelity
spectroscopic imaging. The most recent imaging-spectroscopy study of a
solar flare has shown that MeerKAT has already met, at least
partially, several of the key solar flare requirements. Benefiting
from its excellent $uv$-coverage and high sensitivity, the
observations achieved high-fidelity imaging across the L band,
revealing spatially distinct bright coherent sources (Figure
~\ref{fig:meerkat_fine}), incoherent emission from hot coronal plasma,
and even fainter diffuse components that remain invisible in AIA
observations. The resulting images exhibit both high spatial resolution 
and good overall quality across the entire solar disk, 
demonstrating that MeerKAT can effectively image the full-Sun 
environment while resolving fine coronal structures.
These results represent an advance compared with previous-generation
radio arrays, establishing MeerKAT as a proof of concept for the
next-generation SKA-Mid. Building on MeerKAT's demonstrated
capabilities, SKA-Mid will take solar flare diagnostics with radio
imaging spectroscopy to a fundamentally new level.
With improved solutions for handling solar signal attenuation, 
SKA-Mid is expected to achieve imaging fidelity beyond 
that of the current generation of instruments.
Crucially, this unprecedented image quality, combined
with the potential for high temporal cadence, 
will enable flexible
observing modes optimized for flare studies.  

\section{Solar observations and further progress}

Radio observations provide powerful diagnostics of these processes
through multiple emission mechanisms, each sensitive to
flare-associated non-thermal electrons and their evolving coronal
environments. Over the past decades, the development of imaging
spectroscopy has led to revolutionary advances in flare studies
\citep{2011SSRv..159..225W}. Current-generation interferometric arrays
image each time-frequency pixel for both polarization products,
allowing us to investigate spatially resolved sources and their
temporal and spectral evolution in unprecedented detail, and to
connect the observed emission structures with the underlying physical
processes that govern energy release, particle acceleration, and
transport in the corona.

\begin{figure}[htb!]
    \centering
    \includegraphics[width=0.99\linewidth]{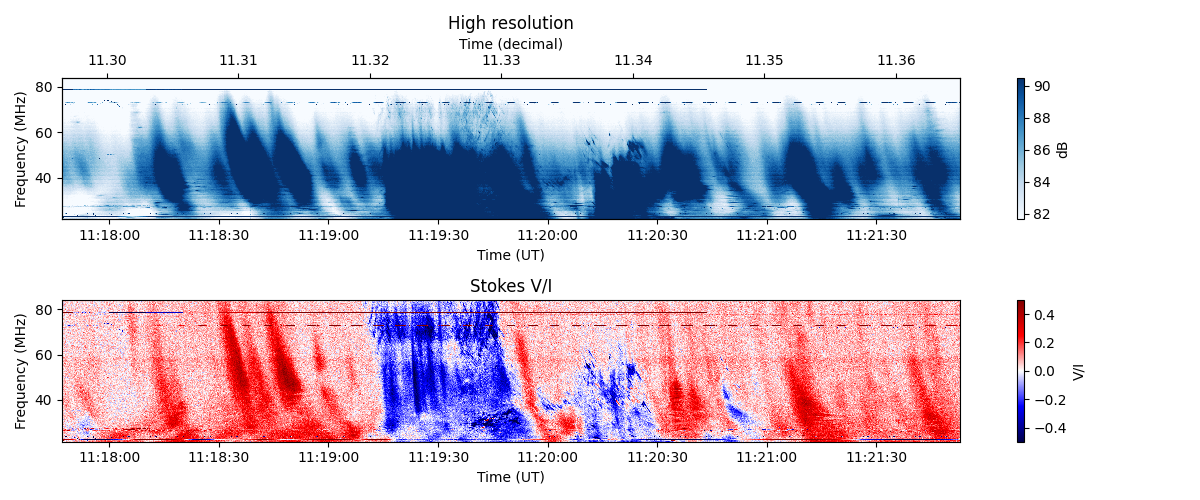}
    \caption{ \textit{Top:} Dynamic spectrum of solar radio emission
    observed by NenuFAR on 2024 July 11 (Sun-integrated, in relative units).
    \textit{Bottom:} Polarization of the same time interval showing
    reverse of polarization around 11:19:30 UT. The spectra are from
    \href{NenuFAR}{https://nenufar.obs-nancay.fr}.}
    \label{fig:July11_2024}
\end{figure}

However, to achieve further progress in flare diagnostics, several key
limitations of current radio arrays must be addressed by future
instruments. First, the available temporal resolution - even at the
millisecond scale is often insufficient to resolve the shortest-lived
coherent bursts \citep{2026ApJ...998L..46L}, whose rapid evolution
contains critical information about the transport of energetic
electrons and their interaction with the surrounding plasma. 
Second, high spatial resolution combined with a sufficiently wide 
field of view is required, so that fine coronal structures 
can be resolved
while simultaneously covering the entire solar disk and the overlying
corona extending to at least several solar radii. Third, improved
$uv$-coverage is essential for producing high-fidelity synthesis
images that enable the investigation of multiple spatially distinct
and spectrally different sources evolving simultaneously during
flares. In addition, broad and continuous frequency coverage, together
with accurate full-polarization calibration, is crucial for
distinguishing between different emission mechanisms and for deriving
quantitative diagnostics of the energetic electron population and the
coronal magnetic environment. Figure \ref{fig:July11_2024} shows that
while Stokes I do not display much difference, the polarization signal
indicates that the conditions of emission completely changed. Finally,
solar-dedicated calibration schemes and observing operations are also
essential to ensure stable performance during high-intensity solar
observations. 

For example, beam-formed modes may deliberately trade a modest amount
of image fidelity for substantially improved temporal and frequency
resolutions. Given SKA intrinsically high imaging performance
(Figure~\ref{fig:meerkat_fine}), such a trade-off is expected to be
highly advantageous for flare diagnostics \citep{2026ApJ...998L..46L}.
The inclusion of the outer antennas will further enhance the
$uv$-coverage, providing much better spatial resolution for resolving
fine coronal structures and improved sensitivity to faint, extended
features. Together, these advances will enable investigations of radio
sources at previously unattainable temporal and spatial resolution.
The SKA will open new opportunities for advances in key science
areas-particularly particle acceleration and transport, 
the associated magnetic reconnection and energy release processes 
at the shorter scales than available before, and the subsequent
energy transformation and coupling between non-thermal electrons 
and the heated plasma.

Another tantalizing opportunity stems from the interplanetary bursts,
where variations in type III radio burst drift rates could be caused
by magnetic-field direction changes, not just density inhomogeneities
\citep{2026ApJ...999..134C}. Many observables like drift-rate
reductions, delays, and fine structure to be observed with the SKA
could be natural consequences of beam propagation along disturbed
magnetic fields, especially magnetic switchback-like structures, the
topic poorly understood in solar physics. 

\section{Summary}

Frequency-time-resolved imaging observations of solar radio bursts has
brought a qualitatively new look at solar phenomena. The solar radio
burst images are found to change at sub-second scales: the centroid
locations at a given frequency are moving often with superluminal
speeds, and the angular extent of the radio sources at a given
frequency is also evolving with time. Often, the images are likely to
be due to scattering, but the variation of image size from one burst
type to another suggests that we do not fully understand all processes
responsible for the source size variations. 

Evidently the observations at selected frequencies are unable to shed
light on the origin of the fine time-frequency structure of solar
radio emission. The lack or absence of imaging information often makes
it impossible to understand the origin of the fine structures. 

A handful of observations able to resolve time-frequency structures
strongly suggests the diagnostic potential of radio emission from 50
MHz to 10 GHz range is not fully explored. The SKA will improve our
understanding of the fine structure of solar radio bursts, by
providing much needed imaging information on the locations and sizes
of the sources over a broad range of frequencies.  
As all solar corona sources are embedded into magnetized turbulent
plasma, the escape of radio emission is an important consideration to
make. 

\section{Acknowledgements}
This work is partially supported by the Leverhulme Trust 
(Research Fellowship RF-2025-357) and STFC/UKRI grant ST/Y001834/1.
We acknowledge support from the International Space Science 
Institute for the LOFAR team \url{http://www.issibern.ch/teams/lofar/}.
This work has made use of NASA's 
Astrophysics Data System Bibliographic Services.
N. Chrysaphi was supported by the Space It Up! project, 
funded by the Italian Space Agency (ASI) and the Ministry 
of University and Research (MUR), 
under contract No. 2024-5-E.0--CUP I53D24000060005.

\bibliographystyle{abbrvnat-maxbibnames4}
\bibliography{chapter} 

\end{document}